\documentstyle[times]{mn}
\input{epsf}

\newcommand{\bez}{\begin{eqnarray*}}
\newcommand{\eez}{\end{eqnarray*}}
\newcommand{\be}{\begin{equation}}
\newcommand{\ee}{\end{equation}}
\newcommand{\beq}{\begin{eqnarray}}
\newcommand{\eeq}{\end{eqnarray}}
\newcommand{\bc}{\begin{center}}
\newcommand{\ec}{\end{center}}

\newbox\grsign \setbox\grsign=\hbox{$>$} \newdimen\grdimen \grdimen=\ht\grsign
\newbox\simlessbox \newbox\simgreatbox \newbox\simpropbox
\setbox\simgreatbox=\hbox{\raise.5ex\hbox{$>$}\llap
     {\lower.5ex\hbox{$\sim$}}}\ht1=\grdimen\dp1=0pt
\setbox\simlessbox=\hbox{\raise.5ex\hbox{$<$}\llap
     {\lower.5ex\hbox{$\sim$}}}\ht2=\grdimen\dp2=0pt
\setbox\simpropbox=\hbox{\raise.5ex\hbox{$\propto$}\llap
     {\lower.5ex\hbox{$\sim$}}}\ht2=\grdimen\dp2=0pt
\def\simgt{\mathrel{\copy\simgreatbox}}
\def\simlt{\mathrel{\copy\simlessbox}}

\def\sT{\sigma_{\rm T}}
\def\TC{T_{\rm C}}
\def\RC{R_{\rm C}}
\def\tC{t_{\rm C}}
\def\LE{L_{\rm E}}
\def\dd{{\rm d}}
\def\ctg{{\rm ctg}}
\def\dM{\dot{M}}

\def\dMt{\dot{M}_{\rm tot}}

\def\thi{\theta_\infty}
\def\vpi{\varphi_\infty}
\def\thob{\theta_{\rm obs}}
\def\vpob{\varphi_{\rm obs}}

\def\lmax{l_{\rm max}}

\begin{document}

\title[Emitting caustics in wind-fed X-ray sources]
{Free-fall accretion and emitting caustics in wind-fed X-ray sources}

\author[Andrei F. Illarionov and Andrei M. Beloborodov]
{\parbox[]{6.8in}{Andrei F. Illarionov$^{1,2}$ and 
Andrei M. Beloborodov$^{1,2}$}\\
$^1$Stockholm Observatory, SE-133 36 Saltsj\"obaden, Sweden\\
$^2$Astro-Space Center of Lebedev Physical Institute,
  Profsoyuznaya 84/32, 117810 Moscow, Russia}

\date{Accepted, Received}

\maketitle


\begin{abstract}
In wind-fed X-ray binaries the accreting matter is Compton cooled and falls 
freely onto the compact object. The matter 
has a modest angular momentum $l$ and accretion is quasi-spherical at large 
distances from the compact object. Initially small non-radial velocities grow 
in the 
converging supersonic flow and become substantial in the vicinity of the 
accretor. The streamlines with $l>(GMR_*)^{1/2}$ (where $M$ and $R_*$ are the 
mass and radius of the compact object) intersect outside $R_*$ and form 
a two-dimensional caustic which emits X-rays.
The streamlines with low angular momentum, $l<(GMR_*)^{1/2}$, run into the 
accretor. 
If the accretor is a neutron star, a large X-ray luminosity results.
We show that the distribution of accretion rate/luminosity over the 
star surface is sensitive to the angular momentum distribution of the accreting
matter.
The apparent luminosity depends on the side from which the star is observed 
and can change periodically with the orbital phase of the binary.
The accretor then appears as a `Moon-like' X-ray source.

\end{abstract}




\section{Introduction}

Wind-fed accretion is believed to occur in massive X-ray binaries 
(see e.g. King 1995 for a review). The massive donor companion (OB star) 
produces a substantial wind, up to $\dot{M}_w\sim 10^{-6}{\rm M}_\odot/$yr, 
which is partly captured by the compact companion.
The wind material is captured from an accretion cylinder of radius 
\beq
R_a=\frac{2GM}{w^2}
\eeq
where $w$ is the wind velocity and $M$ is the mass of the 
accretor (Hoyle \& Lyttleton 1939, Bondi \& Hoyle 1944). The typical 
$w\approx 10^8$~cm~s$^{-1}$ for OB stars, so that 
$R_a\approx 3\times 10^{10}(M/{\rm M}_\odot)(w/10^8)^{-2}$~cm.
If the wind is isotropic then the accretion rate is 
$\dot{M}\approx (1/4)(R_a/A)^2\dot{M}_w $ where $A$ is the binary separation. 
The accretion rate is substantial in close binaries only, with orbital periods 
$P\sim$ a few days. The captured fraction, $\dot{M}/\dot{M}_w$, can be 
increased if the donor is a Be star which has a prominent slow equatorial wind 
(see e.g. van Paradijs \& McClintock 1995) or if the wind is prefocused by 
the tidal effects (Blondin, Stevens \& Kallman 1991).

Gas captured from the accretion cylinder falls many decades in radius down to 
the radius of the compact object $R_*$ where the X-rays are produced. $R_*$ 
equals $r_g=2GM/c^2\approx 3\times 10^5(M/{\rm M}_\odot)$~cm for a black hole 
(BH) and 
about $3r_g$ for a neutron star (NS). Owing to orbital rotation of the binary, 
the captured gas possesses a net angular momentum with respect to the accretor. 
The average angular momentum $\bar{\bf l}$ can be estimated 
(Illarionov \& Sunyaev 1975; Shapiro \& Lightman 1976). It is directed 
perpendicularly to the binary plane and equals $\bar{l}_z=\zeta(1/4)\Omega R_a^2$
where $\Omega=2\pi/P$ is the angular velocity of the binary  and the numerical 
factor $\zeta\simlt 1$ depends on adopted assumptions (see e.g. Wang 1981; 
Livio et al. 1986; Ruffert 1997, 1999). The angular momentum is small 
and the infall is radial at $R\gg R_*$.

Compton cooling by the central X-ray source makes the 
inflow highly super-sonic inside the Compton radius, 
$\RC\sim 10^{10}{\rm cm}< R_a$ (see Illarionov \& Kompaneets 1990 and 
Section 2.2). Initially 
small non-radial velocities $v_\perp=l/R$ grow in the freely falling flow
and exceed the radial velocity component $v_r\sim (GM/R)^{1/2}$ at 
$R_d\sim l^2/GM$. Accretion can be assumed to be radial if $R_d\ll R_*$, i.e.
if $l\ll l_*$ where
\beq
   l_*=(GMR_*)^{1/2}.
\eeq
The deviations from the radial pattern are 
important if $l$ is comparable to $l_*$.

In BH and NS binaries, $l_*\sim r_g c$ and 
$\bar{l}_z/r_g c\approx 1.5 \zeta P^{-1}(M/{\rm M}_\odot)(w/10^8)^{-4}$
where the binary period $P$ is measured in days. 
The observed $P$ in massive (OB) X-ray binaries is typically a few 
days (White, Nagase \& Parmar 1995; Tanaka \& Lewin 1995).
One thus concludes that $\bar{l}_z$ is about $l_*$ in these 
systems. Note that the angular momentum of a particular streamline  
varies substantially around the average value.
For instance, if the flow is in solid body rotation at 
$R\gg R_*$ then $l$ is highest for streamlines in the equatorial plane and 
vanishes on the polar axis. 

Matter with $l<l_*$ runs directly into the accretor before it reaches
the equatorial plane. If the accretor is a neutron star then 
a strong shock results and X-rays are produced 
(Zel'dovich \& Shakura 1969; Shapiro \& Salpeter 1975). 
The surface brightness of the star is determined by the distribution of 
the accretion rate over its surface, $\dd\dM/\dd S$.
In this paper, we find that $\dd\dM/\dd S$
is sensitive to the angular momentum distribution in the flow.
We assume a weakly magnetised NS ($B\simlt 10^{8}$~G),
so that the magnetic field does not affect the ballistic
trajectories of the freely falling matter.
The resulting surface brightness of the star is inhomogeneous and
the apparent luminosity depends on the side from which the star is 
observed. The apparent luminosity can then change as the binary executes 
its orbital period (we dub it the `Moon' effect). 

By contrast, if the accretor is a black hole then matter with 
$l<l_*$ plunges into the event horizon without producing substantial 
emission.

The streamlines with $l>l_*$ intersect in the equatorial plane (the plane of
symmetry) at $R>R_*$. The loci of the intersections form a 
two-dimensional caustic. If the accretor is a black hole then the caustic is 
the only source of X-rays from the accretion flow.

The paper is organised as follows. In Section~2 we briefly review the 
pattern of wind-fed accretion on large scales, at distances $\sim R_a$ from
the accretor. In Section 3 we write down the equations of the freely falling 
flow inside the Compton radius. In Sections 4 and 5 we focus on the very 
vicinity of the compact object. We discuss asymmetric accretion onto the 
surface of a NS (Section~4) and then caustics outside the accretor (Section~5).


\section{Wind-fed accretion on large scales}

\subsection{The trapping of the wind matter}

The radius of the accretion cylinder is small compared to the binary 
separation, 
$R_a\simlt 10^{11}$ cm $\ll A\sim 10^{12}$ cm, and hence the flow in the 
cylinder is nearly plane-parallel before it gets trapped by the gravitational 
field of the accretor. The wind reaches the compact companion on a time-scale 
$A/w\sim 10^4$~s which is much shorter than the orbital period $P\sim 10^6$~s.
As a first approximation, one can assume the accretor to be at rest and
the flow to be axisymmetric around the line connecting the two companions. 

Let us introduce coordinates $(x,y,z)$ so that the $y$-axis is directed 
from the donor to the accretor and the $z$-axis is perpendicular to the 
binary plane, and choose the coordinate origin at the location of the accretor
(see Fig.~1). Each streamline of the flow is specified by two impact 
parameters $(x_0,z_0)$ at $R\gg R_a$. 
The initial angular momentum of a streamline $(x_0,z_0)$ is 
${\bf l}=(-z_0w,0,x_0w)$ and its absolute value equals $l=bw$ where 
$b=(x_0^2+z_0^2)^{1/2}$. The net ${\bf l}$ integrated over a ring $(b,b+db)$ 
vanishes, which is a consequence of the assumed symmetry around the $y$-axis.

\begin{figure}
\begin{center}
\leavevmode
\epsfxsize=8.4cm
\epsfbox{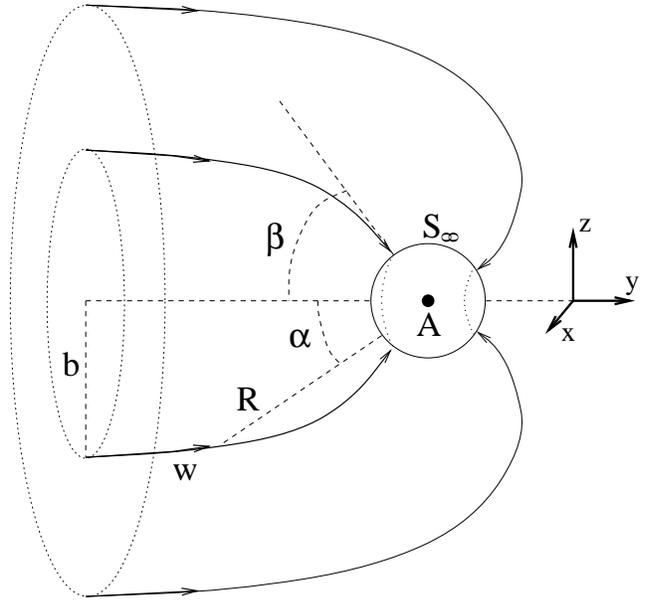}
\end{center}
\caption{ Schematic picture of wind-fed accretion. The initially 
plane-parallel flow is captured by the accretor and transformed into a 
spherical inflow. $S_\infty$ denotes a sphere of radius $R\sim\RC$ (eq. 7).
Inside $S_\infty$, the flow is falling freely (Section 3).
The coordinate origin is at the location of the accretor, $A$.
}
\end{figure}

For an initially super-sonic wind, a bow shock forms at distance $\sim R_a$
from the accretor. Hunt (1971) first studied gas dynamics behind the shock
and showed that a spherically symmetric inflow forms at $R<R_a$ (see also
Petrich et al. 1989; Ruffert 1997, 1999). 
Fig.~1 shows the picture of accretion.
The transformation of the uniform plane-parallel flow into the isotropic spherical 
infall can be described as follows.
A streamline with an initial impact parameter $b$ eventually infalls
radially at some angle $\beta$ with respect to the $-y$ axis,
\beq
\nonumber
  \left(\frac{b}{R_a}\right)^2=\frac{1-\cos\beta}{2}, \qquad b^2=x_0^2+z_0^2.
\eeq
This transformation induces a map 
$(x_0,z_0)\rightarrow (\theta,\varphi)$, 
\be
  \frac{x_0}{R_a}=\frac{\sin\theta\sin\varphi}
{\sqrt{2(1+\sin\theta\cos\varphi)}}, \quad
  \frac{z_0}{R_a}=\frac{\cos\theta}
{\sqrt{2(1+\sin\theta\cos\varphi)}}.
\ee
Here $\theta$ is the polar angle 
measured from the $z$-axis and $\varphi$ is the azimuthal angle measured in 
the $xy$ plane from the ($-y$)-axis. 
Note that the boundary of the accretion cylinder
$x_0^2+z_0^2=R_a^2$ transforms into one point $\theta=\pi/2$, $\varphi=\pi$.

The mapping (3) assumes that the flow is laminar behind the bow shock.
Numerical simulations show that the flow is unstable 
if the bow shock is strong, with a high Mach number.
However, in the case of modest Mach numbers, the fluctuations are weak and
the flow is approximately laminar (e.g. Blondin et al. 1990; Ruffert 1997, 1999). 
This is the most likely case if accretion occurs in the radiation field of a 
luminous X-ray source (see below). Note also that 
the streamlines cross the shock nearly normally (Hunt 1971); therefore
the shock does not generates vorticity in the flow (Landau \& Lifshitz 1987).

\subsection{Compton heating/cooling}

If the central X-ray source is luminous, $L>10^{-3}\LE$ where 
$\LE=4\pi cGMm_p/\sT$ is the Eddington luminosity, then Compton 
heating/cooling affects strongly the dynamics of accretion (Ostriker et al. 
1976; Illarionov \& Kompaneets 1990). We here discuss two effects:
(i) On large scales, $R\simgt R_a$, the wind matter is preheated and its 
Mach number is reduced to ${\cal M}\sim 1$. (ii) On small scales, $R<\RC$
(see eq.~7), the accreting matter is cooled by the X-rays and falls 
super-sonically (Zel'dovich \& Shakura 1969). 

\subsubsection{Preheating of the wind}

The initial temperature of the wind is $T_0\sim 10^5$~K and the typical 
Mach number is ${\cal M}_0=w/c_s\sim 10$ where $c_s=(10kT/3m_p)^{1/2}$ is the 
sound speed. One can evaluate how ${\cal M}$ decreases along a streamline
when approaching $R\sim R_a$. The Compton heating/cooling is dominant in the
energy balance of the highly ionised plasma in the accretion cylinder (see
e.g. Igumenshchev, Illarionov \& Kompaneets 1993). Then the energy balance 
reads
\be
  \frac{\dd T}{\dd t}=-\frac{2}{3}T{\rm div}{\bf v}
                      +\frac{T_{\rm C}-T}{t_{\rm C}}
\ee
where $T$ and ${\bf v}$ are the plasma temperature and velocity, respectively,
$T_{\rm C}\sim 10^8$~K is the Compton temperature of the X-ray source, 
$t_{\rm C}=3\pi m_ec^2 R^2/\sT L$ is the time-scale for Compton cooling, 
and $L$ is the source luminosity. Since $\TC\gg T$ at $R\simgt R_a$, the 
Compton heating term is dominant and we keep only this term on the right-hand 
side of the energy equation, 
\be
   \frac{\dd T}{\dd t}\approx\frac{T_{\rm C}}{t_{\rm C}}. 
\ee

In the upstream region, the wind matter is falling freely. 
From the angular momentum conservation we have
\beq
\nonumber
  R^2\frac{\dd\alpha}{\dd t}=bw
\eeq
where $b$ is the impact parameter of the streamline and $\alpha$ is the angle 
between the radius vector ${\bf R}$ and the $(-y)$ axis (see Fig.~1).
Combining this equation with the energy equation~(5), we get 
\beq
\nonumber
  \frac{\dd T}{\dd\alpha}=\frac{\TC}{\tC}\frac{R^2}{bw}
=\frac{\sT L\TC}{3\pi m_ec^2bw}=const.
\eeq
Taking $R=b/\sin\alpha\approx b/\alpha$, we get the temperature
\beq
  T\approx T(R)=\frac{\sT L\TC}{3\pi m_ec^2Rw}.
\eeq
At $R=R_a$ it gives a Mach number
\begin{eqnarray}
\nonumber
 {\cal M}^2=\frac{3m_pw^2}{10kT}
           \approx \frac{w}{10^8}
                   \left(\frac{L}{0.1\LE}\right)^{-1}
                   \left(\frac{\TC}{10^{8}}\right)^{-1}. 
\end{eqnarray}
As a result of the Compton preheating, ${\cal M}$ decreases markedly below 
the initial ${\cal M}_0$, and, correspondingly, the strength of the bow shock 
is reduced. In bright hard sources, the heated wind may become subsonic at 
$R\simgt R_a$ and then the shock disappears.

\subsubsection{Compton radius}

Compton heating leads to an inflow-outflow pattern of accretion with 
${\cal M}\sim 1$ down to the Compton radius (Illarionov \& Kompaneets 1990; 
Igumenshchev, Illarionov, \& Kompaneets 1993),
\be
\RC=\frac{GMm_p}{5k\TC}\approx 
   3\times 10^{9}\left(\frac{M}{{\rm M}_\odot}\right)
             \left(\frac{\TC}{10^8}\right)^{-1} {\rm cm}. 
\ee
Typically, $\RC$ is $\sim 10$ times smaller than $R_a$. 

Inside $\RC$, a spherical inflow forms. Its temperature exceeds $\TC$, so that
the gas is cooled by the X-rays rather than heated (see eq.~4).
The cooling leads to a high Mach number of the spherical inflow, ${\cal M}\gg 1$.

\subsection{The trapped angular momentum}

When the accretor orbital motion is taken into account, the flow pattern
is no longer symmetric around the $y-$axis and
the captured matter should have a small net angular momentum 
$\bar{l}_z\sim \Omega R_a^2$ with respect to the accretor.
One would like to know the distribution of $l$ around $\bar{l}_z$ in the 
accretion flow. This distribution governs the flow dynamics 
in the vicinity of the accretor where
deviations from the spherical pattern become substantial. 

Assuming a weak bow shock, the flow is almost laminar all the way.
Then the transformation of the accretion cylinder into the spherical inflow 
is given by equation~(3). The corresponding transformation 
of angular momentum is not known. The analysis of small perturbations in 
a converging flow (e.g. Lai \& Goldreich 2000 and references therein) shows
that the rotational mode, $v_{\rm rot}\propto r^{-1}$, has the fastest growth 
inwards, while sonic modes are damped. The rotational mode is probably dominant
on $\RC$ and we restrict our consideration to this mode and the associated 
angular momentum.

A particular streamline with impact parameters $(x_0,z_0)$ contributes
to the net trapped angular momentum proportionally to $\Omega x_0^2$
(see e.g. Shapiro \& Lightman 1976). 
If this magnitude conserves along a streamline then mapping (3) also 
determines the distribution of the trapped $l$ over angles $\theta,\varphi$ 
at $R\ll\RC$,
\be
  l_z(\theta,\varphi)=l_0\frac{\sin^2\theta\sin^2\varphi}
                              {1+\sin\theta\cos\varphi},
   \qquad \bar{l}_z=\frac{l_0}{2}.
\ee
We assume that the infall angular momentum 
is associated with rotation around the $z$-axis only, i.e. we assume 
$v_\varphi\gg v_\theta$. Then the orbital angular momentum of a streamline is
\beq
  l=\frac{l_z}{\sin\theta}.
\eeq
The corresponding non-radial velocity is $v_\perp=v_\varphi=l_z/R$.
The distribution (8) is {\it not} axisymmetric, which leads to an 
essentially three-dimensional pattern of accretion in the vicinity of the 
compact object. 

In the other limiting case, the trapped $l$ is efficiently redistributed 
between the streamlines so that they come to solid body rotation around the
$z$-axis with a common angular velocity $\omega$. Then
\be
  l_z(\theta)=l_0\sin^2\theta, \qquad \bar{l}_z=\frac{2}{3}l_0.
\ee
The non-radial velocity is $v_\perp=v_\varphi=\omega R\sin\theta=l_z/R$ and
the infall is axisymmetric.

\begin{figure}
\begin{center}
\leavevmode
\epsfxsize=9.4cm
\epsfbox{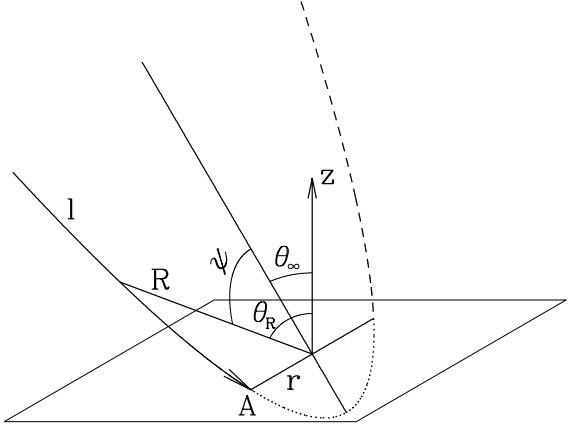}
\end{center}
\caption{ Geometry of the free fall. At point A (radius $r=(l/l_*)^2R_*$), the 
ballistic trajectory intersects with the symmetric trajectory from below the 
equatorial plane.
If $r<R_*$ then the trajectory instead runs into the accretor, before it
can reach the equatorial plane. 
}
\end{figure}

\section{Free fall}

In this section, we write down the equations describing the matter free fall 
in Newtonian gravity (the relativistic equations are given in an accompanying 
paper, Beloborodov \& Illarionov 2000). We consider sub-Eddington sources only,
where radiation pressure does not affect the flow dynamics.
We are interested in the part of 
the ballistic trajectories before the collision with the accretor or the 
intersection with the equatorial plane, the symmetry plane of the flow.

At $R\ll\RC$, the free fall is nearly parabolic. 
A parabolic trajectory is described by the equation
\be
  R(\psi)=\frac{l^2R_*}{l_*^2(1-\cos\psi)}, \qquad l_*=\sqrt{GMR_*}
\ee
where $\bf l$ is the orbital angular momentum of a streamline and 
$\psi$ is the angle between the changing radius-vector of the streamline 
$\bf R$ and its initial radius-vector at infinity $\bf R_\infty$ 
(see Fig.~2). The label `$\infty$' corresponds to distances $R\sim\RC$. 

The angular distribution of the accretion rate at a sphere $S_R$ of radius $R$
is different from the initial uniform distribution at $S_\infty$:
the rotation defocuses the inflow and makes it non-uniform. 
A streamline that starts at $\thi,\vpi$ will cross $S_R$ 
at $\theta,\varphi$ which satisfy the relations (see Fig.~2)
\be
  \cos\theta=\cos\theta_\infty\cos\psi,
\ee
\be
  \sin(\varphi-\varphi_\infty)=\frac{\sin\psi}{\sin\theta},
\ee
where $\cos\psi=1-l^2R_*/l_*^2R$ (from eq.~11).
The accretion rate
distribution at $S_R$ is determined by the Jacobian of the mapping
$(\theta_\infty,\varphi_\infty)\rightarrow (\theta,\varphi)$.
After some algebra we get the Jacobian
\be
  \Delta=\frac{\partial(\cos\theta,\varphi)}
                   {\partial(\cos\theta_\infty,\varphi_\infty)}=
  \cos\psi -\ctg\theta_\infty\frac{\partial\cos\psi}{\partial\theta_\infty}
  -\frac{\ctg\psi}{\sin\theta_\infty}
       \frac{\partial\cos\psi}{\partial\varphi_\infty}
\ee
where
\beq
\nonumber
  \frac{\partial\cos\psi}{\partial\thi}=-\frac{2lR_*}{l_*^2R}\;\frac{\partial l}
                                         {\partial\thi}, \qquad
  \frac{\partial\cos\psi}{\partial\vpi}=-\frac{2lR_*}{l_*^2R}\;
                                         \frac{\partial l}{\partial\vpi}.
\eeq
The mapping is one-to-one if $\Delta>0$ at any $\thi,\vpi$. Vanishing of the 
Jacobian implies intersection of the ballistic trajectories. The 
streamlines then approach each other and the pressure effects must switch on 
and prevent the streamlines from intersecting. If the velocity of the 
approaching is super-sonic then shocks must occur. 

It is convenient to rewrite equation~(14) as
\be
  \Delta=\left(1-\lambda^2\right)\left(1+q\frac{\partial\lambda}{\partial\vpi}
\right)+\ctg\thi\frac{\partial\lambda^2}{\partial\thi}
\ee
where $\lambda=(l/l_*)(R_*/R)^{1/2}$ and 
$q=2(2-\lambda^2)^{-1/2}\sin^{-1}\thi$.
In the axisymmetric case ($l$ does not depend on $\vpi$)
the Jacobian is positive if $\dd l/\dd\sin\theta_\infty>0$, i.e. if 
$l(\theta_\infty)$ increases towards the equatorial plane. Then the ballistic  
approximation can be used down to the surface of the accretor 
(in the case $l<l_*$) or down to the equatorial plane 
(in the case $l>l_*$). The condition $\dd l/\dd\sin\theta_\infty>0$ 
is satisfied for e.g. $l$-distribution~(10). 
 
With $\Delta>0$ we have a simple expression for the angular distribution of 
the accretion rate $\dM$ over a sphere of radius $R$
\be
  \frac{\dd\dM}{\dd \Omega}=\frac{1}{\Delta}\frac{\dd\dM}{\dd\Omega_\infty},
  \qquad \dd\Omega=\dd\cos\theta\dd\varphi.
\ee


\section{Neutron star as a Moon-like X-ray source}

In this section, we study inflows with sufficiently small angular 
momentum, $l<l_*$, which have no caustics outside the accretor. 
Each ballistic streamline runs into the accretor. 
Throughout this section we will assume that the accretor is a (weakly 
magnetised) neutron star. The collision of accreting matter with the star is 
accompanied by a strong shock. We assume that the shock is radiatively 
efficient and that it is held down to the star, i.e. the height of the 
shock is small compared to $R_*$. This situation is likely to take place at 
sufficiently high accretion rates (see Shapiro \& Salpeter 1975).
Then the accreting matter is in free fall until it reaches the surface of the 
star.

\subsection{Surface distribution of $\dM$}

The presence of non-zero angular momentum leads to an inhomogeneous 
distribution of $\dM$ over the surface of the star. 
We now compute this distribution.
With a homogeneous accretion rate through $S_\infty$,
$\dd\dM/\dd\Omega_\infty=\dMt/4\pi$, equation~(16) yields
\be
  \frac{\dd\dM}{\dd S}(\theta,\varphi)=\frac{\dMt}{4\pi R_*^2\Delta}.
\ee

First consider the axisymmetric case and for illustration take 
the $l$-distribution~(10)
which corresponds to solid body rotation at $R\rightarrow\infty$. 
The accretion rate $\dd\dM/\dd S$ is then given by
\beq
\nonumber
 \frac{\dd\dM}{\dd S}(\theta)=\frac{\dMt}{4\pi R_*^2[1+(l_0/l_*)^2
                                                       (3\cos^2\thi-1)]}.
\eeq
The flux of matter impinging the accretor increases towards the 
equatorial plane (Fig.~3a). This is a consequence of the fact that $\sin\theta$
increases along a rotating trajectory, so that the flow gets concentrated 
towards $\theta=\pi/2$. In the example shown in Fig.~3a the 
accretion rate at the equator is twice as large as that at the polar cap.

\begin{figure}
\epsfxsize=8.0cm
\epsfysize=8.0cm
\epsfbox{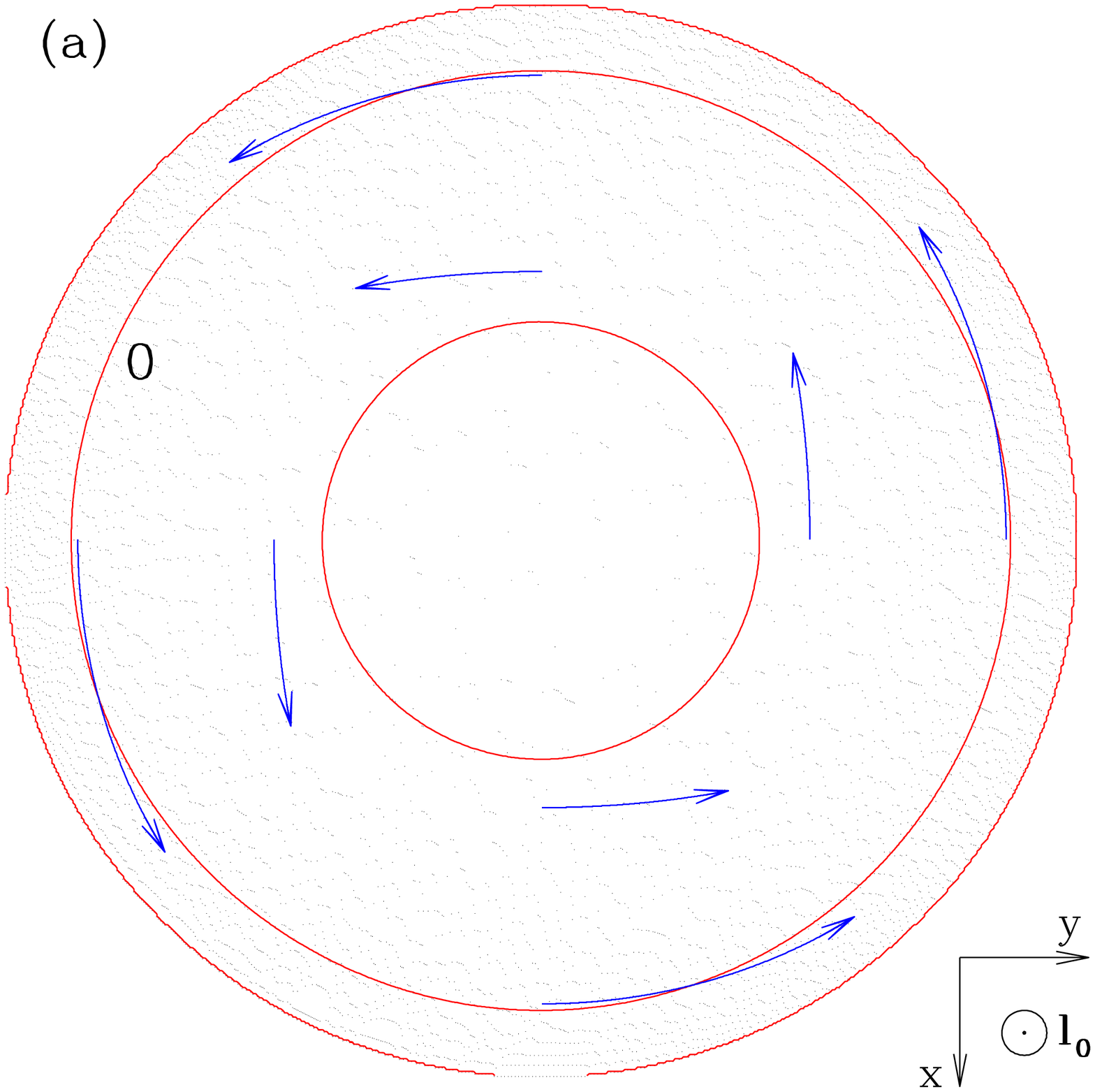}
\caption{The distribution of $\dd\dM/\dd S$ over the surface of the accretor.
Dark regions correspond to high $\dd\dM/\dd S$. Plots (a-c) show the view from 
the top (the accretion flow is symmetric about the equatorial plane). 
The angular momentum of the flow is given by equation~(18) with 
$l_0=0.5(GMR_*)^{1/2}$. Plots (a), (b), and (c) show the cases
$a=0$, $a=0.5$, and $a=1$, respectively; (d) is same as (c) but the moving
star is viewed from the head-on direction in the equatorial plane.
Solid curves are the contours of constant $\dd\dM/\dd S$ with a 
logarithmic step 0.15 in plots (a) and (b), and 0.3 in plots (c) and (d).
The label 0 marks contours $\dd\dM/\dd S=\dMt/4\pi$ ($\Delta=1$).
The arrows show the projections of free-fall streamlines onto the star surface 
for $\thi=\pi/6,\pi/3$ and $\vpi=0,\pi/2,\pi,3\pi/2$ (see eq.~12,13).
}
\end{figure}
\begin{figure}
\epsfxsize=8.0cm
\epsfysize=8.0cm
\epsfbox{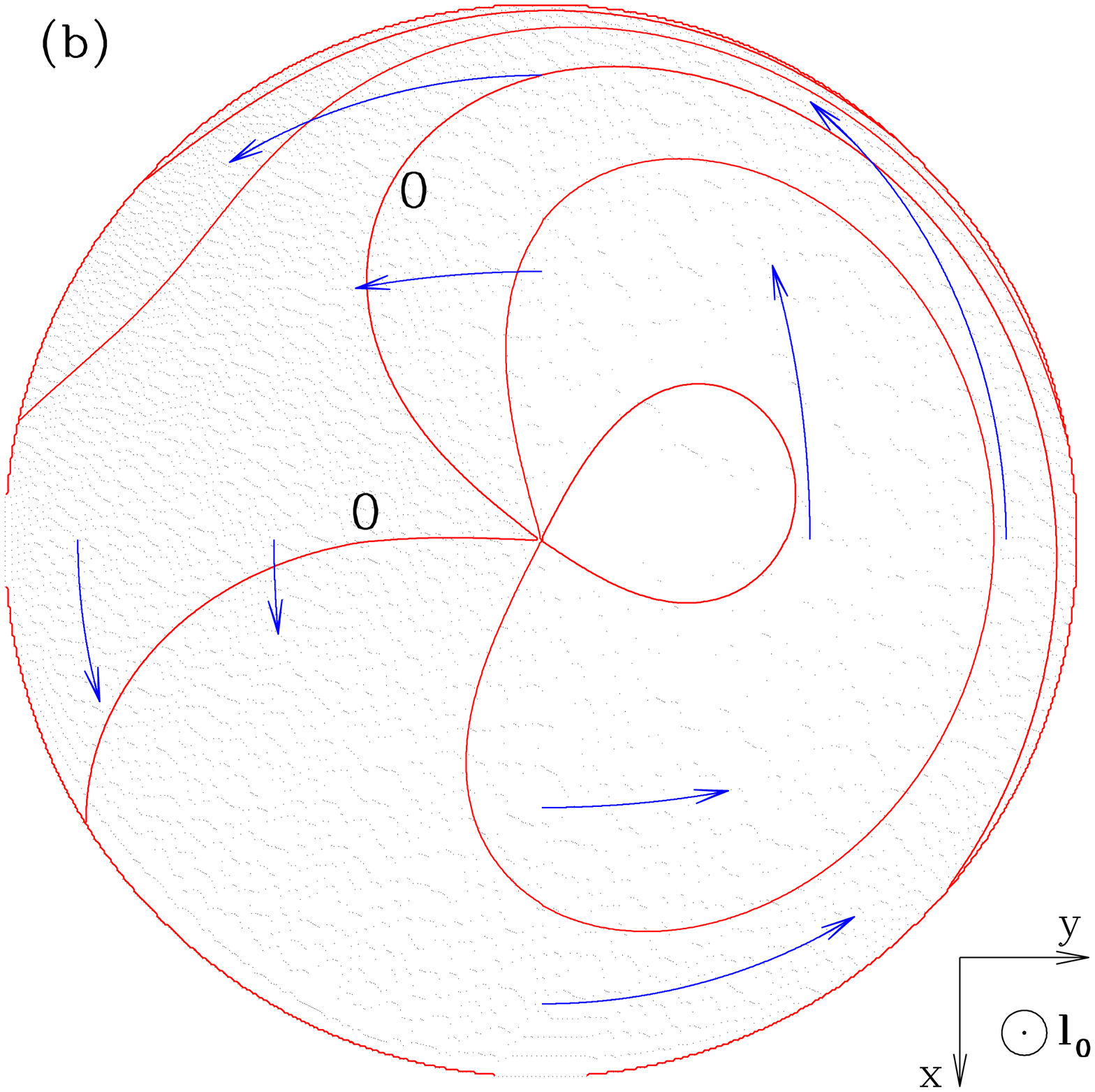}
\epsfxsize=8.0cm
\epsfysize=8.0cm
\epsfbox{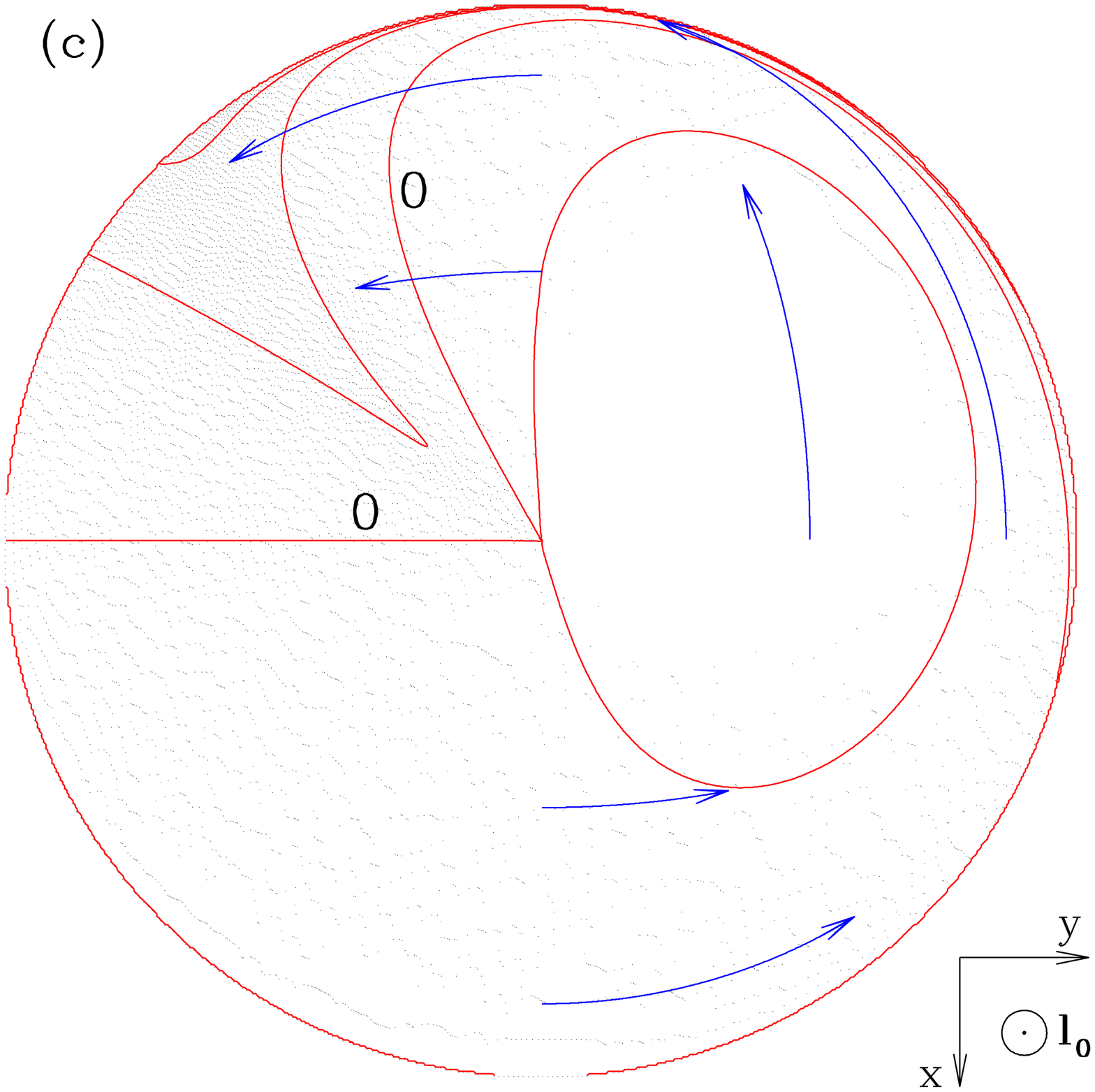}
\epsfxsize=8.0cm
\epsfysize=8.0cm
\epsfbox{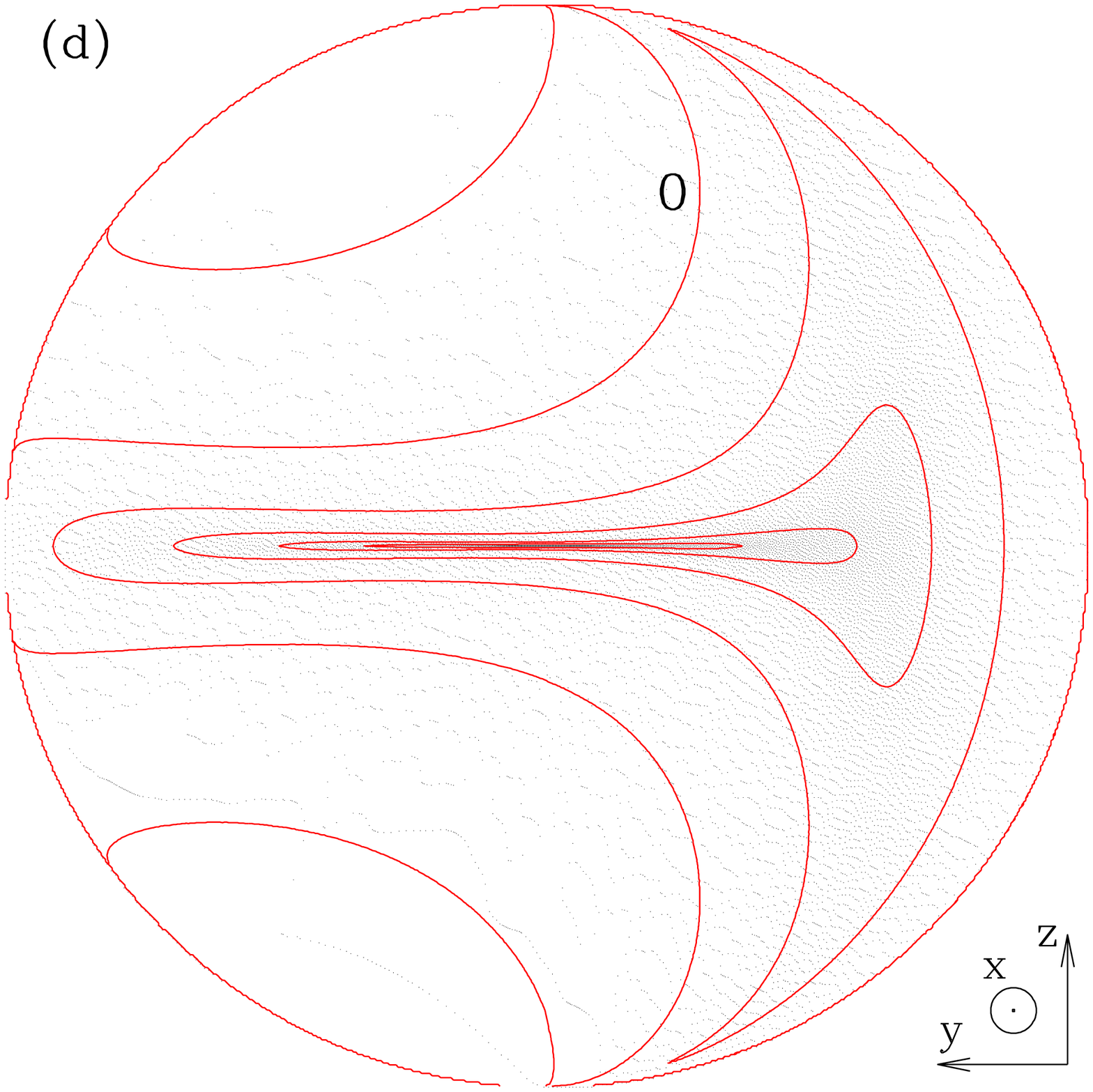}
\end{figure}

In general the inflow does not need to be axisymmetric. In particular, 
the trapped $l$ in the `tail' ($\varphi_\infty\sim \pi$) may differ from 
$l$ trapped at $\varphi_\infty\sim 0$. For example, the distribution (8)
gives the whole set of $\varphi$-harmonics, $m=1,2,...$. To study the effects 
of the $\varphi$-asymmetry, we consider here the first ($m=1$) mode, i.e.
\be
  l(\thi,\vpi)=l_0\sin\thi(1-a\cos\vpi), \qquad a\leq 1.
\ee
The condition that $l<l_*$ over $S_\infty$ requires $l_0(1+a)<l_*$.
Under this condition, equation~(15) yields $\Delta>0$ at $R>R_*$, i.e. 
there are no intersections of the ballistic trajectories outside the accretor.

We have computed $\dd\dM/\dd S$ on the surface of the star for inflows with 
$l_0=0.5l_*$ and three values of $a=0$, $0.5$, and $1$ (Fig.~3).
At $a\neq 0$ the accretion rate is no longer axisymmetric. The inhomogeneity 
of $\dd\dM/\dd S$ increases with increasing $a$. One can see two effects:

\begin{enumerate}

\item
The accretion flow concentrates towards the equatorial plane along
the streamlines with the highest angular momentum (the maximum $l=\lmax$ is at 
$\thi=\pi/2$ and $\vpi=\pi$). For $l_0=0.5l_*$ and $a=1$ we have $\lmax=l_*$ 
and the corresponding streamline marginally touches the star at $\theta=\pi/2$
and $\varphi=3\pi/2$. Here $\Delta=0$ and $\dd\dM/\dd S\rightarrow\infty$ 
(Fig.~3cd). This critical point is a `seed' of the outward caustic: further 
increase in $l_0$ would result in $\lmax>l_*$ and then the streamlines collide
in the equatorial plane outside the star (see Section~5 and Fig.~5).

\item
The flow concentrates around the meridian $\varphi\sim 7\pi/4$ producing a 
`Moon-like' spot on the accretor. The spot is formed by the trajectories 
starting 
at $\vpi\sim 3\pi/2$ where $\partial l/\partial \vpi$ reaches its minimum
$-al_0\sin\thi$ (see eq.~18). Angular momentum determines the velocity
of rotation in the $\varphi-$direction. Like usual one-dimensional motion
with position-dependent velocity, the flow with $\partial l/\partial \vpi>0$
diverges in the $\varphi-$direction ($\dd\dM/\dd\varphi$ decreases) while
the flow with $\partial l/\partial \vpi<0$ converges ($\dd\dM/\dd\varphi$
decreases). Therefore $\dd\dM/\dd S$ peaks at the trajectories with
minimum $\partial l/\partial \vpi$. These trajectories start at $\vpi=3\pi/2$
with angular momentum $l_0\sin\thi$ and they collide with the star at 
$\sim 7\pi/4$.

\end{enumerate}

The Moon effect is clearly seen from a simple analytical consideration.
Assuming a small $l_0$ and neglecting the $\lambda^2$ terms in equations (13,15)
one gets
\beq
\nonumber
 \varphi=\vpi+\frac{l_0}{l_*}\sqrt{2}\left(1-a\cos\vpi\right),\\
\nonumber
 \Delta=1+a\frac{l_0}{l_*}\sqrt{2}\sin\vpi.\hspace*{1.3cm}
\eeq
The dependences $\varphi(\vpi)$ and $\Delta(\vpi)$ describe a cycloid 
$\Delta(\varphi)$ in its standard parametric form. The peak of the accretion rate 
$\dd\dM/\dd S=(\dMt/4\pi R_*^2)[1-a(l_0/l_*)\sqrt{2}]^{-1}$ is achieved at the 
meridian $\varphi=3\pi/2+(l_0/l_*)\sqrt{2}$. In the case $l_0=0.5l_*$ it yields 
$\varphi=270^o+ 40^o$.

\begin{figure}
\epsfxsize=8.2cm
\epsfbox{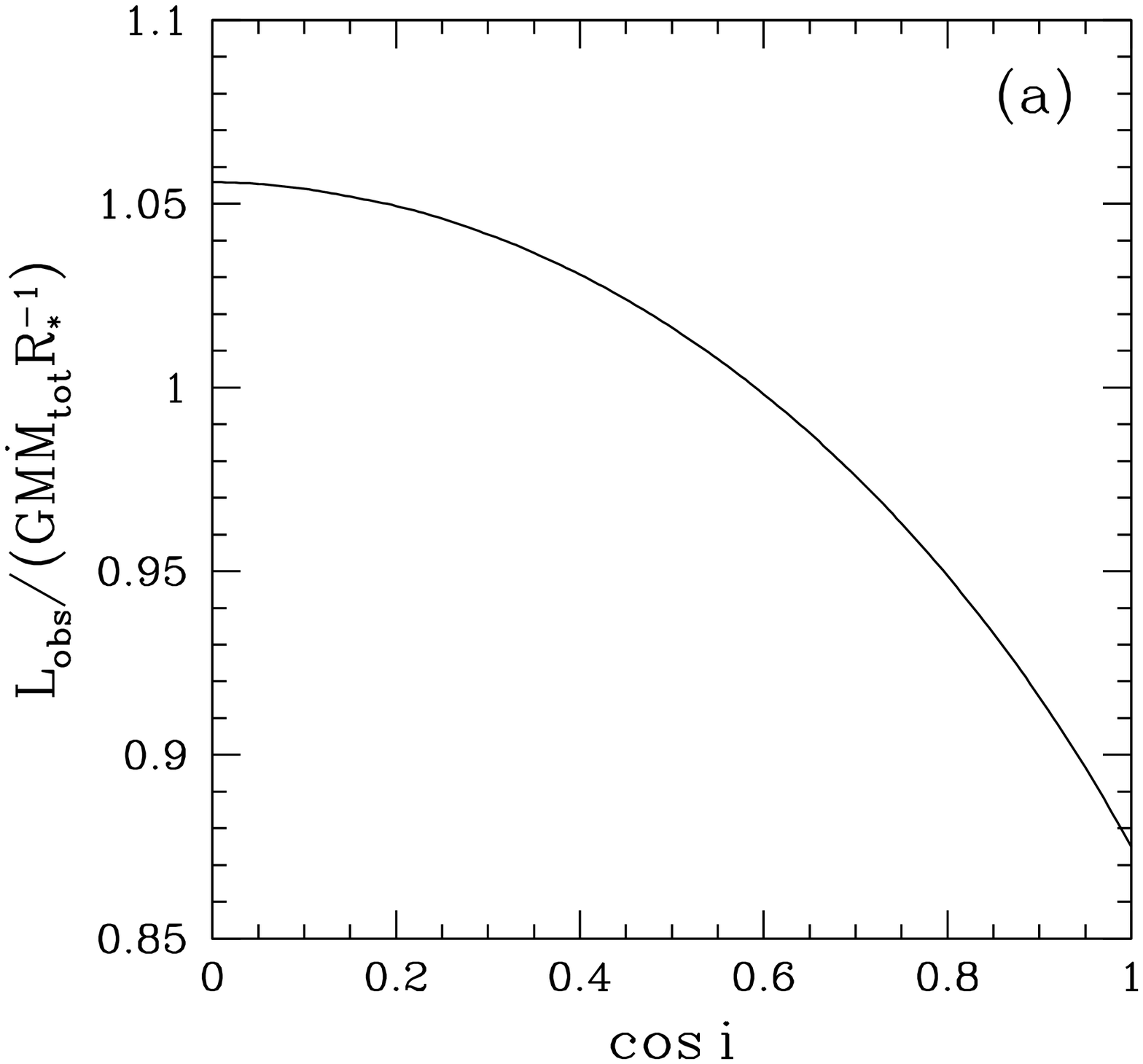}
\caption{ The apparent luminosity $L_{\rm obs}$ normalised by 
the total true luminosity $L=GM\dMt/R_*$.
(a) -- $L_{\rm obs}$ as a function of the binary inclination $i$ in the 
case of $a=0$, $l_0=0.5l_*$ (see eq.~18). In this case, 
$L_{\rm obs}$ does not depend on the orbital phase of the binary $\phi$ 
owing to the axial symmetry of the accretion flow. 
(b) -- $L_{\rm obs}$ as a function of $\phi$ in the case of $a=0.5$, 
$l_0=0.5l_*$. The solid, long-dashed, short-dashed, and dotted curves 
correspond to inclinations $i=\pi/2,\pi/3,\pi/6$, and $0$, respectively. 
(c) -- Same as (b) but with $a=1$. 
}
\end{figure}
\begin{figure}
\epsfxsize=8.2cm
\epsfbox{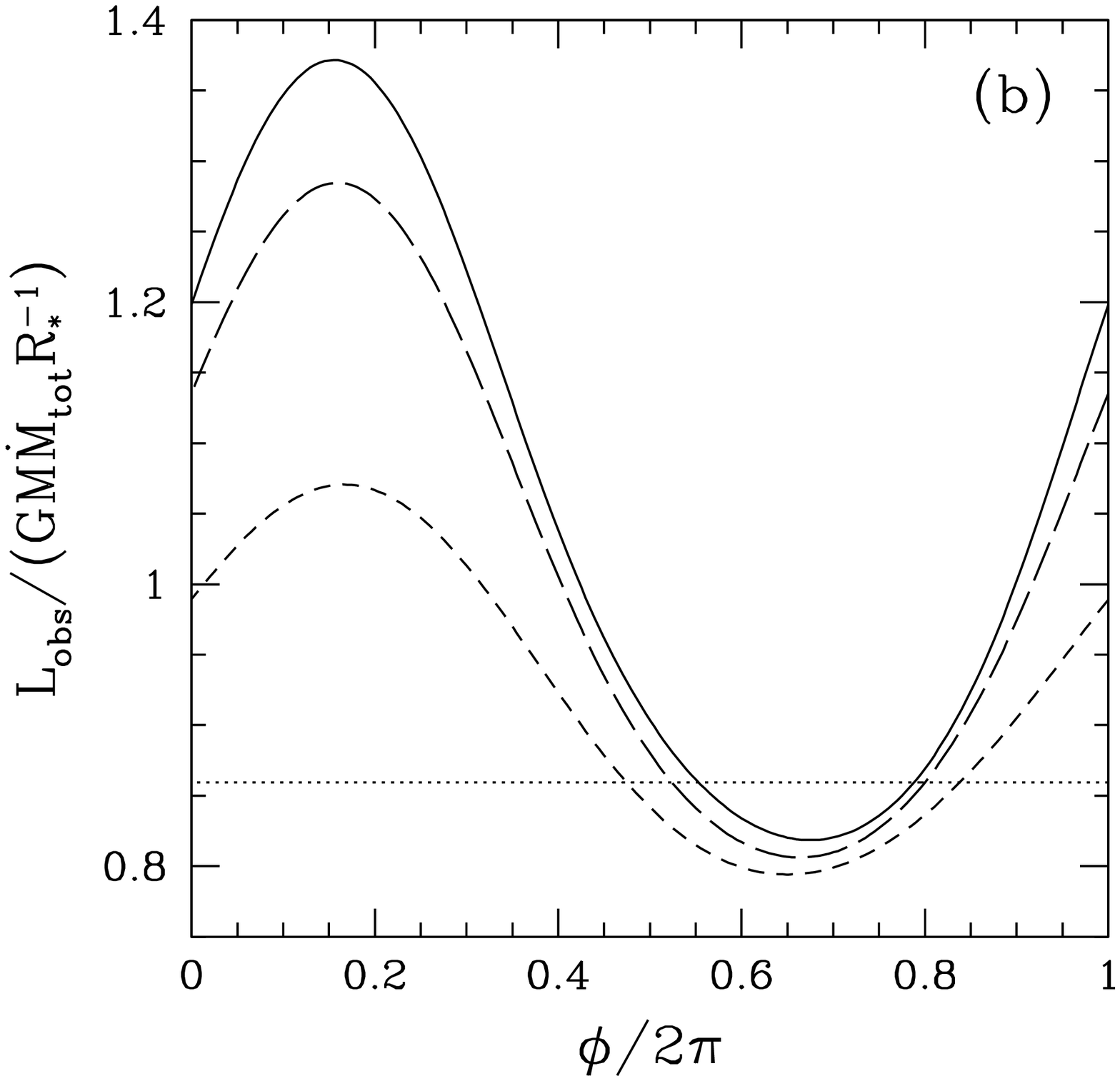}
\epsfxsize=8.2cm
\epsfbox{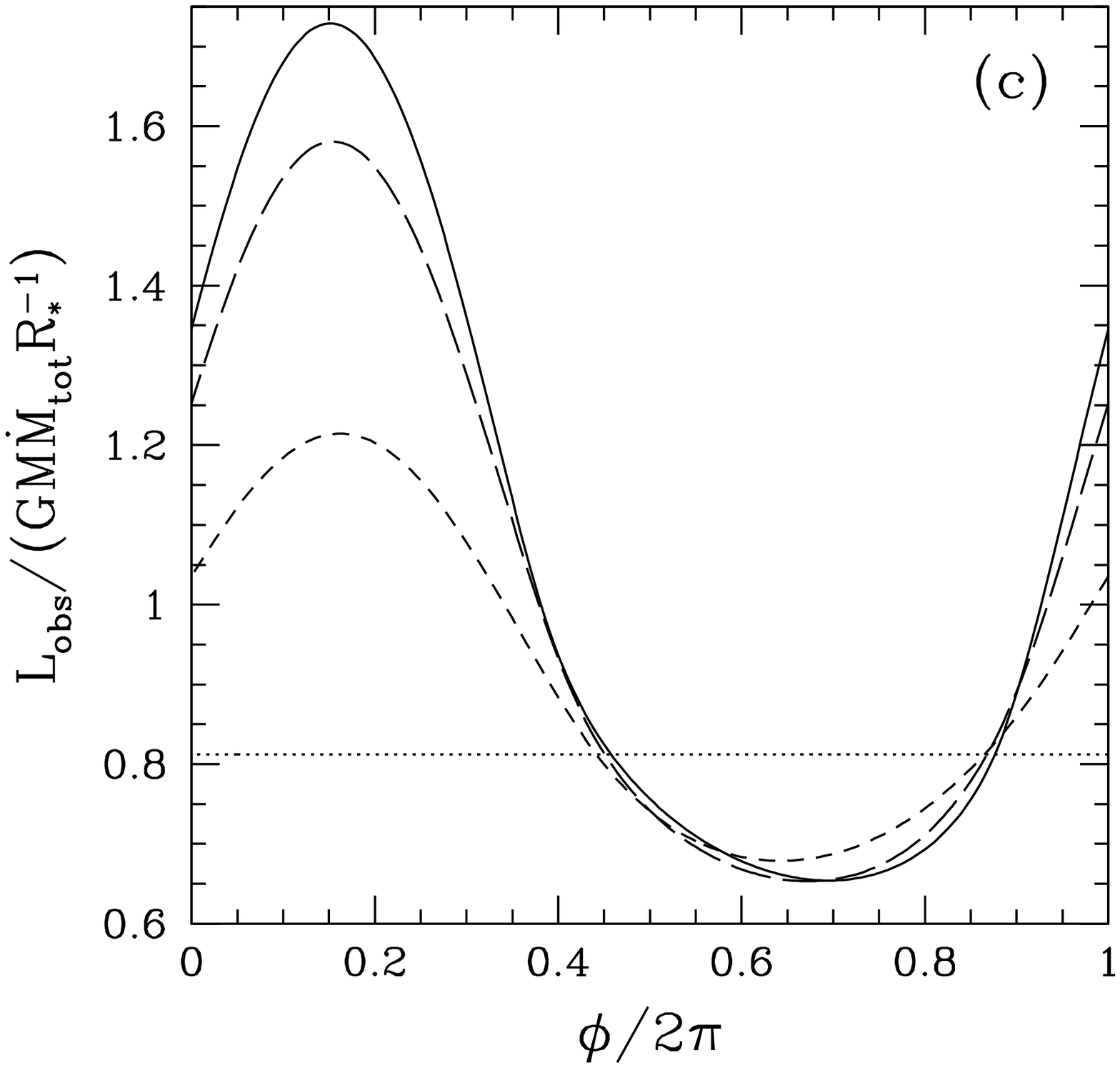}
\end{figure}

\subsection{The observed luminosity}

On the NS surface, the flux of accreting mass is converted into 
radiation with efficiency $GM/R_*c^2$. The emerging radiation flux is given by
\be
  F(\theta,\varphi)=\frac{GM}{R_*}\frac{\dd\dM}{\dd S}.
\ee
The distribution $\dd\dM/\dd S$ shown in Fig.~3 also displays the brightness
distribution as seen by an observer located at the polar axis (if one assumes 
an isotropic intensity of the produced radiation). 

The apparent luminosity of the accreting star depends on the binary 
inclination $i$ and the orbital phase $\phi$. We choose $\phi=0$ at the 
superior conjunction of the accretor, i.e. when the companion is 
between the observer and the accretor. Then the angular position of the 
observer as viewed from the accretor is given by $\thob=i$ and $\vpob=-\phi$.

Let ${\bf\Omega}$ and ${\bf\Omega}_{\rm obs}$
be unit vectors corresponding to $(\theta,\varphi)$ and $(\thob,\vpob)$,
respectively, and $\mu=({\bf\Omega\cdot\Omega}_{\rm obs})$. The apparent 
luminosity is 
\beq
\nonumber
  L_{\rm obs}(i,\phi)=4\int F(\theta,\varphi)\,\mu H(\mu)\,\dd S\\
             =\frac{GM\dMt}{\pi R_*}\int \mu H(\mu)\,\dd\Omega_\infty.
\eeq
Here $H(\mu)$ is the Heaviside step function. To compute the integral,
we substitute $\mu=\cos\theta\cos i+\sin\theta\sin i\cos(\varphi+\phi)$ and
take $\theta(R_*)$ and $\varphi(R_*)$ from equations (12,13).

Fig.~4 shows $L_{\rm obs}$ found for accretion flows with 
$l_0=0.5l_*$ and $a=0$, $0.5$, and $1$ (same cases as shown in 
Fig.~3). One sees strong variations of $L_{\rm obs}$ with the orbital phase. 
The amplitude of the variability reaches $\sim 3$ at $i=\pi/2$ and vanishes
at $i=0$. The maximum at $\phi\sim 7\pi/4$ is produced by 
the bright spot on the surface of the star (see Fig.~3). Note that at large
$i$ one should also take into account the eclipse by the donor.


\section{Caustics outside the accretor}

\subsection{Formation of caustics}

We now address the case $l>l_*$. If the accretion flow is symmetric 
about the equatorial plane then a streamline with 
$l>l_*$ coming from above will collide in this plane with the 
symmetric streamline coming from below. The collision occurs at 
$r=(l/l_*)^2R_*$.
We study inflows with $\Delta>0$ (see eq.~14), so that there are no 
intersections of the ballistic trajectories outside the equatorial 
plane\footnote{In general, depending on the distribution 
$l(\theta_\infty,\varphi_\infty)$, such `early' intersections 
are possible. In that case,
the increased pressure near $\Delta=0$ would alter the trajectories. It
would cause a relatively modest focusing effect on the streamlines, without 
substantial energy release. By contrast, the eventual collision in the 
symmetry plane liberates a large fraction of the infall kinetic energy.}. 

The collision in the equatorial plane is associated with a couple of shocks 
that envelope the caustic from above and below. The caustic shock is similar 
to the shock on the surface of a NS (now the symmetry plane plays the role of 
a `hard surface'). Like the case of collision with the star, we assume that the 
shocks are pinned to the caustic. The matter is then in free fall until 
it reaches the equatorial plane. The flux of mass impinging the caustic 
determines its brightness.

Note that the occurrence of caustics and the associated energy release are
{\it not} caused by the centrifugal barrier often mentioned in the literature. 
The radial infall is not stopped by rotation when matter reaches the caustic.
Rather, the two symmetric streams `miss' the center and collide.
As a result they cancel $\theta-$components of velocity and continue
to accrete in the equatorial plane. The subsequent accretion proceeds via a fast 
disc (see Beloborodov \& Illarionov 2000). Note that the free fall from 
infinity onto the equatorial plane executes only 1/4 of the full turn around 
the $z$-axis ($\varphi-\vpi=\pi/2$). The centrifugal barrier
would stop the infall at the periastron radius $r_p=(l/l_*)^2(R_*/2)$
after 1/2 of the full turn (see eqs. 11, 13, and Fig.~2). 
The collision thus happens {\it before} the centrifugal barrier. 

In the most general case, the streams from above and below are not symmetric 
about the equatorial plane. The caustic may then be a time-dependent 
warped surface. This situation would be the subject of a separate study.
In this paper, we restrict our consideration to the inflows which are symmetric
about the equatorial plane, but not necessary axisymmetric. Then the general
shape of the caustic is an asymmetric ring in the equatorial plane.

\subsection{The caustic shape}

We use the polar coordinates $(r,\varphi)$ on the equatorial plane.
A streamline starting at $\thi,\vpi$ reaches the plane at 
\beq
r=\frac{l^2(\thi,\vpi)}{GM}, \qquad \varphi=\vpi+\pi/2.
\eeq
We thus have a mapping $(\theta_\infty,\varphi_\infty)\rightarrow (r,\varphi)$.
With a homogeneous accretion rate through 
$S_\infty$, we get the flux of matter impinging the caustic on one side at 
given $r,\varphi$, 
\be
   \frac{\dd\dM}{\dd S}(r,\varphi)
   =\frac{\dMt}{4\pi r}\left|\frac{\partial r}{\partial \cos\thi}\right|^{-1}.
\ee
The equatorial streamlines have the highest angular momentum $l_{\rm max}(\vpi)$ 
and the outer edge of the caustic is defined by the streamlines with 
$\thi\rightarrow \pi/2$,
\beq
  r_0(\varphi)=\frac{l_{\rm max}^2(\vpi)}{GM}, \qquad \vpi=\varphi-\pi/2.
\eeq
If $l(\thi,\vpi)$ is a differentiable function at $\thi=\pi/2$ then 
$\partial l/\partial\cos\thi=0$ and $\dd\dM/\dd S\rightarrow \infty$ at 
$r\rightarrow r_0(\varphi)$, i.e. the flux of matter diverges at the outer edge
of the caustic.

\begin{figure}
\epsfxsize=8.4cm
\epsfysize=8.4cm
\epsfbox{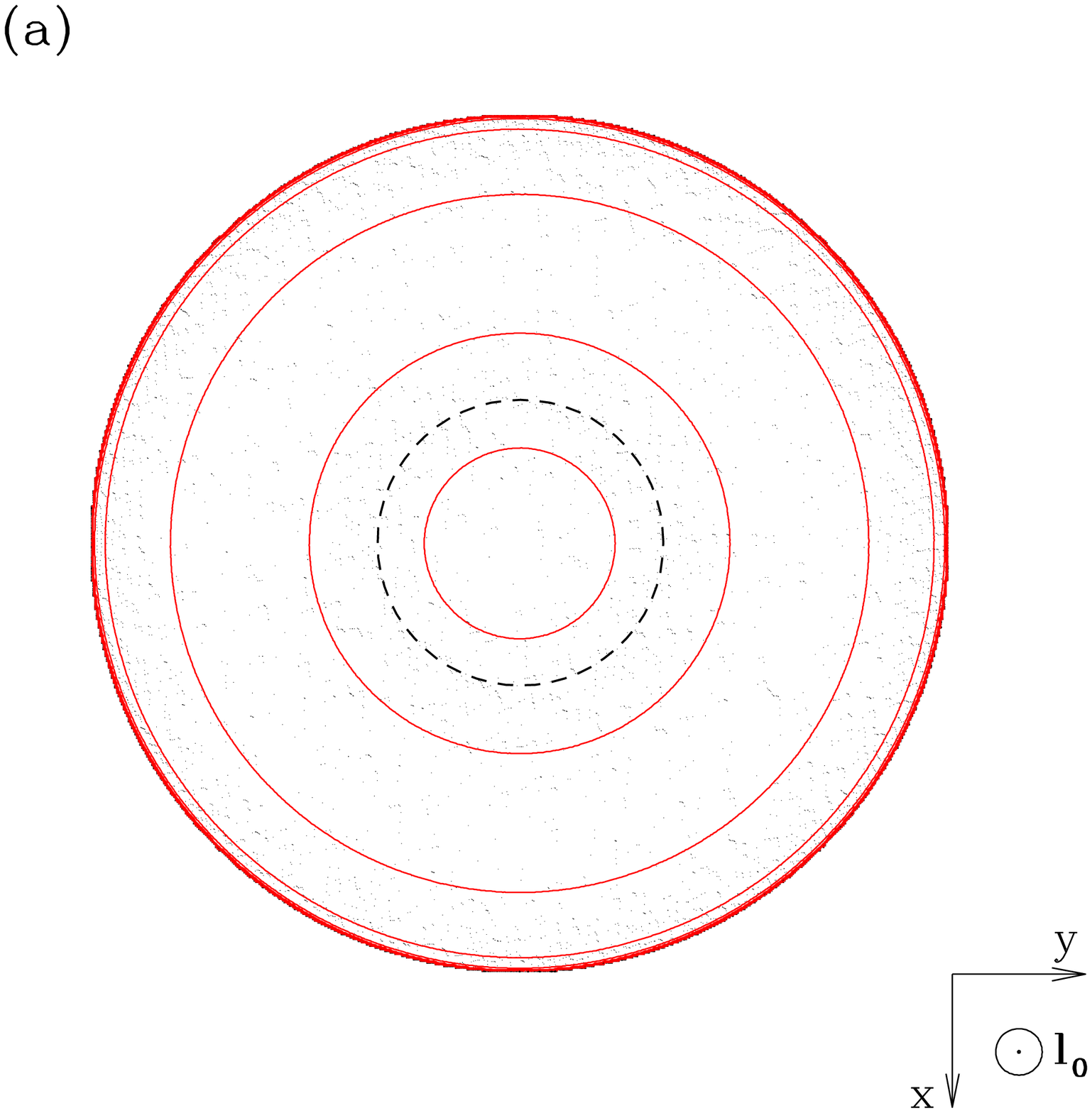}
\end{figure}
\begin{figure}
\epsfxsize=8.0cm
\epsfysize=8.0cm
\epsfbox{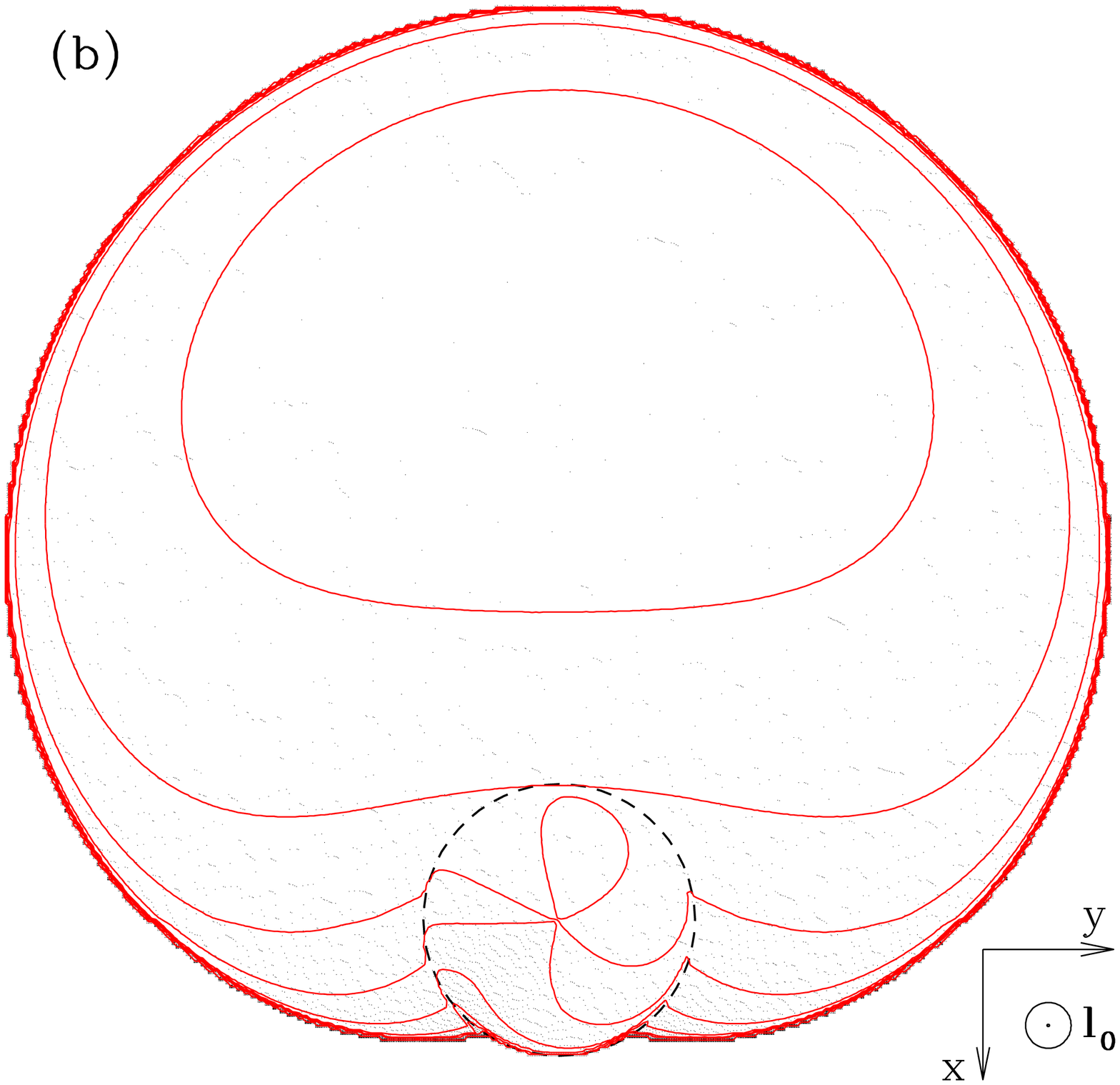}
\caption{ The distribution of $\dd\dM/\dd S$ over the caustic surface 
(at $r>R_*$) and the surface of the accretor (at $r<R_*$). The view is taken
from the top, i.e., the plane of the figure corresponds to the equatorial 
plane, $\theta=\pi/2$. The dashed curve shows the radius $R_*$.  
The solid curves display the contours of constant $\dd\dM/\dd S$ with a 
logarithmic step 0.3. Dark regions correspond to high $\dd\dM/\dd S$.
The assumed angular momentum of the accretion flow is given by equation~(18)
with $l_0=\sqrt{3}l_*$. Plot (a) shows the axisymmetric case ($a=0$) 
and plot (b) shows the case $a=0.5$. 
The outer edge of the caustic is $r_0(\varphi)=3R_*(1-a\sin\varphi)^2$.
}
\end{figure}

As an illustration take the inflow (18). Then
\beq
  \frac{\dd\dM}{\dd S}(r,\varphi)=\frac{\dMt}{8\pi rr_0}
  \left(1-\frac{r}{r_0}\right)^{-1/2}, \quad R_*<r<r_0(\varphi)
\eeq
where $r_0(\varphi)=(l_0^2/GM)(1-a\sin\varphi)^2$.
Note that the $r$-integrated distribution of the accretion rate 
is $\dd\dM/\dd\varphi\approx\dMt/4\pi=const$ at $r_0\gg R_*$.
At a given $\varphi$, the minimum 
$\dd\dM/\dd S=(3\sqrt{3}/8)(\dMt/2\pi r_0^2(\varphi))$ is achieved 
at $r_{\rm min}=(2/3)r_0(\varphi)$. 
In Fig.~5 we take $l_0=\sqrt{3}l_*$ and compare the distributions 
$\dd\dM/\dd S$ in the cases of $a=0$ and $a=0.5$. 

Note two general features of the caustic:

\begin{enumerate}

\item
The asymmetry of $l(\thi,\vpi)$ in the $y$-direction 
results in the asymmetry of the caustic in the $x$-direction.
This is a consequence of the fact that the streamlines execute a 1/4 turn 
by the moment of collision. The caustic thus appears in 
the front of the accretor orbiting the donor star. 
If the trapped $\bar{l}_z$ had the opposite sign, the caustic would 
appear in the rear of the moving accretor.

\item
The outer edge of the caustic is formed by the nearly equatorial streamlines,
$\thi\rightarrow\pi/2$. The edge is sharp, with $\dd\dM/\dd S\rightarrow\infty$
if the $l$-distribution is smooth (differentiable) at $\thi=\pi/2$.

\end{enumerate}


\section{Discussion}

The regime of accretion studied in this paper applies to weakly magnetised 
accretors. In the case of accretion onto a strongly magnetised
neutron star, the effective radius of the accretor is the Alfv\'enic radius, 
$R_A$ and therefore the characteristic $l_*$ in that problem is 
$\sim(GMR_{\rm A})^{1/2}$.

In reality one expects wind-fed accretion flows to be time-dependent.
Numerical simulations of the Compton heated subsonic region at 
$R\simgt\RC$ show that the flow is unsteady (Igumenshchev et al. 1993). 
Fluctuations at $R\simgt\RC$ imply variable boundary conditions 
for the free fall inside $\RC$. The typical time-scale of the variations is of 
order of the free-fall time at $\RC$. The accretion flow is thus expected 
to fluctuate on time-scales $\sim 10$~s.

Throughout the paper we assumed that the flow is symmetric with respect
to the equatorial (binary) plane. If the symmetry is broken,
a warped caustic may form near the accretor. The changes in the 
caustic shape are then driven by the ram pressure of the colliding gas. 
The warped caustic is likely to be unstable, leading to 
oscillations/fluctuations on time-scales $\sim (R_*^3/GM)^{1/2}\sim 0.1$~ms.

The pattern of accretion studied in this paper assumes that the shocks on the
star surface/caustics are radiatively efficient and pinned to the 
star surface and/or the equatorial plane. At a low accretion rate (which 
implies low density) the protons heated in the shock may find it
difficult to pass their energy to the electrons on the free-fall time-scale. 
Then the shocked gas cannot radiate the heat and a variable 
pressure-driven outflow is likely to form. 

On the observational side, the possible modulation of X-ray emission with the
binary period is especially interesting. Note that orbital modulations of soft
X-rays are known to occur in wind-fed systems as a result of photoelectric
absorption in the wind (e.g. Wen et al. 1999; Ba{\l}uci\'nska-Church 
et al. 2000).
The study of orbital modulations in the hard X-ray band would be especially 
helpful since they can be caused by intrinsic anisotropy of the source.


\section*{Acknowledgments}

This work was supported by the Wenner-Gren Foundation for Scientific Research,
the Swedish Natural Science Research Council, and RFBR grant 00-02-16135.

\end{document}